\documentclass[journal]{IEEEtran}
\usepackage[backend=biber, style=ieee]{biblatex}
\AtEveryBibitem{
    \clearfield{urlyear}
    \clearfield{urlmonth}
}
\usepackage{hyperref}

\ifCLASSINFOpdf
  \usepackage[pdftex]{graphicx}
   \DeclareGraphicsExtensions{.pdf,.jpeg,.png}
\else
  \usepackage[dvips]{graphicx}
  \DeclareGraphicsExtensions{.eps}
\fi
\usepackage{amsmath}
\ifCLASSOPTIONcompsoc
  \usepackage[caption=false,font=normalsize,labelfont=sf,textfont=sf]{subfig}
\else
  \usepackage[caption=false,font=footnotesize]{subfig}
\fi
\usepackage[utf8]{inputenc} 
\usepackage[capitalise]{cleveref}
\crefname{equation}{}{}
\usepackage{xcolor}

\begin{document}

\title{Molecular Dynamics Simulations of Mutual Space-Charge Effect between Planar Field Emitters}
%
%
%

\author{Hakon~Valur~Haraldsson, Kristinn~Torfason,~\IEEEmembership{Member,~IEEE,}
        Andrei~Manolescu, and Agust~Valfells~\IEEEmembership{Member,~IEEE.}
\thanks{H.V. Haraldsson, K. Torfason, A. Manolescu, and A. Valfells are with the Department of Engineering, Reykjavík University, Menntavegi 1, IS-102 Reykjavík, Iceland. (Contact Agust Valfells: av@ru.is}}

%
%

\markboth{Journal of \LaTeX\ Class Files,~Vol.~14, No.~8, August~2015}%
{Shell \MakeLowercase{\textit{et al.}}: Bare Demo of IEEEtran.cls for IEEE Journals}
%



\maketitle

\begin{abstract}
Molecular dynamics simulations, with full Coulomb interaction and self-consistent field emission, are used to examine mutual space-charge interactions between beams originating from several emitter areas, in a planar infinite diode. The simulations allow observation of the trajectory of each individual electron through the diode gap. Results show that when the center-to-center spacing between emitters is greater than half of the gap spacing the emitters are essentially independent. For smaller spacing the mutual space-charge effect increases rapidly and should not be discounted. A simple qualitative explanation for this effect is given. 
\end{abstract}

\begin{IEEEkeywords}
Field emission, field emitter arrays, space-charge, diode physics, vacuum electronics, electron sources.
\end{IEEEkeywords}

%
\IEEEpeerreviewmaketitle

\section{Introduction}
%
%
%
%
\IEEEPARstart{F}{ield} emitter arrays (FEAs) are important elements in energy efficient and responsive cathodes for electron beam generation~\cite{whaley_application_2000,dowell_cathode_2010,shiffler_high-current_2004}. Numerous studies have been conducted on FEA design with regard to improving performance. Many of those are concerned with the phenomenon of \textit{screening} or \textit{shielding}, which refers to how the relative placement of field emitting structures affects the local electric field at the emitter tips, hence the current flows~\cite{tang_analysis_2011,harris_dependence_2015,harris_shielding_2015,forbes_screened_2012,forbes_physical_2016,dallagnol_physics-based_2018}. Other studies have looked at how to modify the electric field at the cathode edge to counteract increased emission from the cathode rim~\cite{umstattd_two-dimensional_2001,luginsland_beyond_2002,hegeler_reduction_2002,harris_control_2016}. These previously mentioned studies are primarily concerned with the structure of the applied electric field. It is also necessary to look at space charge effects due to the field emitted current, and how that affects the equilibrium field emission current. This has been done for infinite planar diodes~\cite{barbour_space-charge_1953,lau_electron-emission_1994,luginsland_effects_1996,feng_transition_2006,koh_transition_2006,forbes_exact_2008,rokhlenko_space_2010,darr_unification_2019} and for discrete emitters~\cite{sun_onset_2012,torfason_molecular_2015,jensen_discrete_2015,torfason_molecular_2016}. In FEAs it is necessary to not only to examine how current emitted from a field emitter effects that emitter, but also how it affects emission from neighboring emitters. In this paper we attempt to gain some understanding of the mutual space-charge effects between adjacent emitters. By utilizing planar emitters in the FEA we ensure that there is no screening taking place, giving us the ability to isolate the space charge effect from that of screening.

\section{System Model and Methodology}
\begin{figure}
\centering
\includegraphics[width=0.70\linewidth]{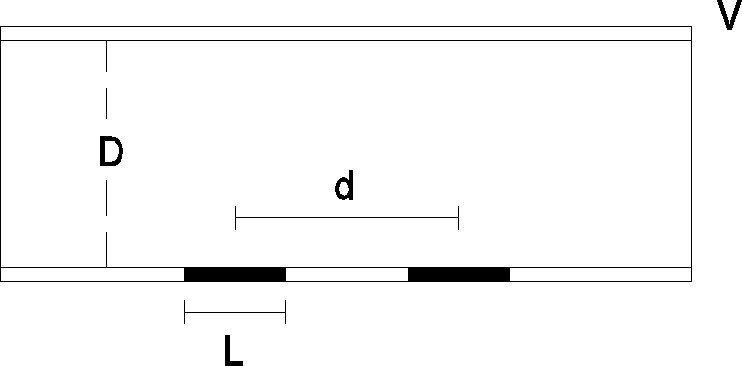}
\caption{Side view of the two emitter system. The system consists of a planar vacuum diode of infinite extent with two square emitting areas of side length $L$. The spacing between the cathode and anode is $D$ and gap voltage is $V$. The spacing between the centers of the emitters is $d$.}
\label{fig:model}
\centering
\vspace{5mm}
\includegraphics[width=0.70\linewidth]{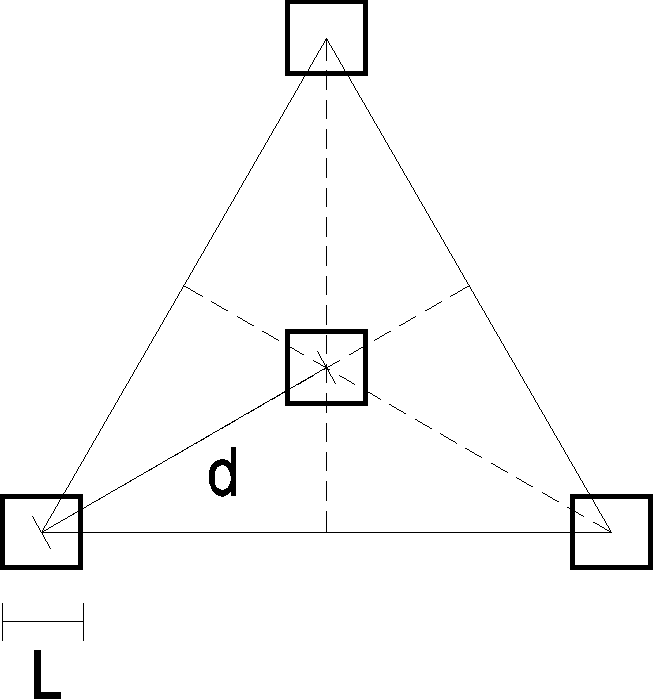}
\caption{Top view of the cathode in the four emitter system. As before, the system consists of a planar vacuum diode, of infinite extent (as shown in Fig.~\ref{fig:model}),
but now with four square emitting areas of side length $L$, 
and with the spacing between the centers of the inner and outer emitters denoted by $d$. The emitters are configured such that the centers of the outer three are on the vertices of an equilateral triangle, while the inner one is at the center of the triangle.}
\label{fig:emitters}
\end{figure}
%
In this paper we study two different types of systems. Firstly, we look at a two emitter system, shown in \cref{fig:model}. This consists of an infinite planar diode with two square emitting patches on the cathode. The side length of each of the patches is $L$, and the center-to-center distance between them is $d$. The diode gap spacing is given by $D$, and the applied voltage is $V$. Secondly, we look at the system shown in \cref{fig:emitters}. This also consists of an infinite planar diode with gap spacing $D$ and gap voltage $V$. However, in this case, there are four square emitters placed so that three of them are centered on the vertices of an equilateral triangle, and the fourth is centered in the middle of the triangle. Each of the square emitting patches has a side length of $L$, but in this case the distance, $d$, denotes the distance from the middle of the triangle to a vertex. The four emitter system has the advantage that it gives the opportunity of distinguishing between interior and exterior emitters.

We use a molecular dynamics approach to modelling field emission and propagation of electrons. The method is the same as has been employed by some of the present authors in a previous paper on field emission from a finite area in a planar diode~\cite{torfason_molecular_2015}. This involves an emission model based on the well known Fowler-Nordheim equation, where the total electric field at the cathode surface is self-consistent with the space-charge and with the image charge induced by the electrons near the cathode. 
The image charge induced in the anode is also included in the model. The Coulomb forces between all pairs of electrons during the propagation through the diode gap are calculated. Hence, our method incorporates the interactions between point-like charges in the system with full resolution. The motion of the electrons is simulated by incrementally displacing each particle, under the action of the total electric field, i.e. the diode field plus the fields corresponding to the space-charge and to the image charge, in time steps of 0.1~fs, using the Verlet method~\cite{torfason_molecular_2015}.

\begin{figure}
        \centering
        \includegraphics[width=0.95\linewidth]{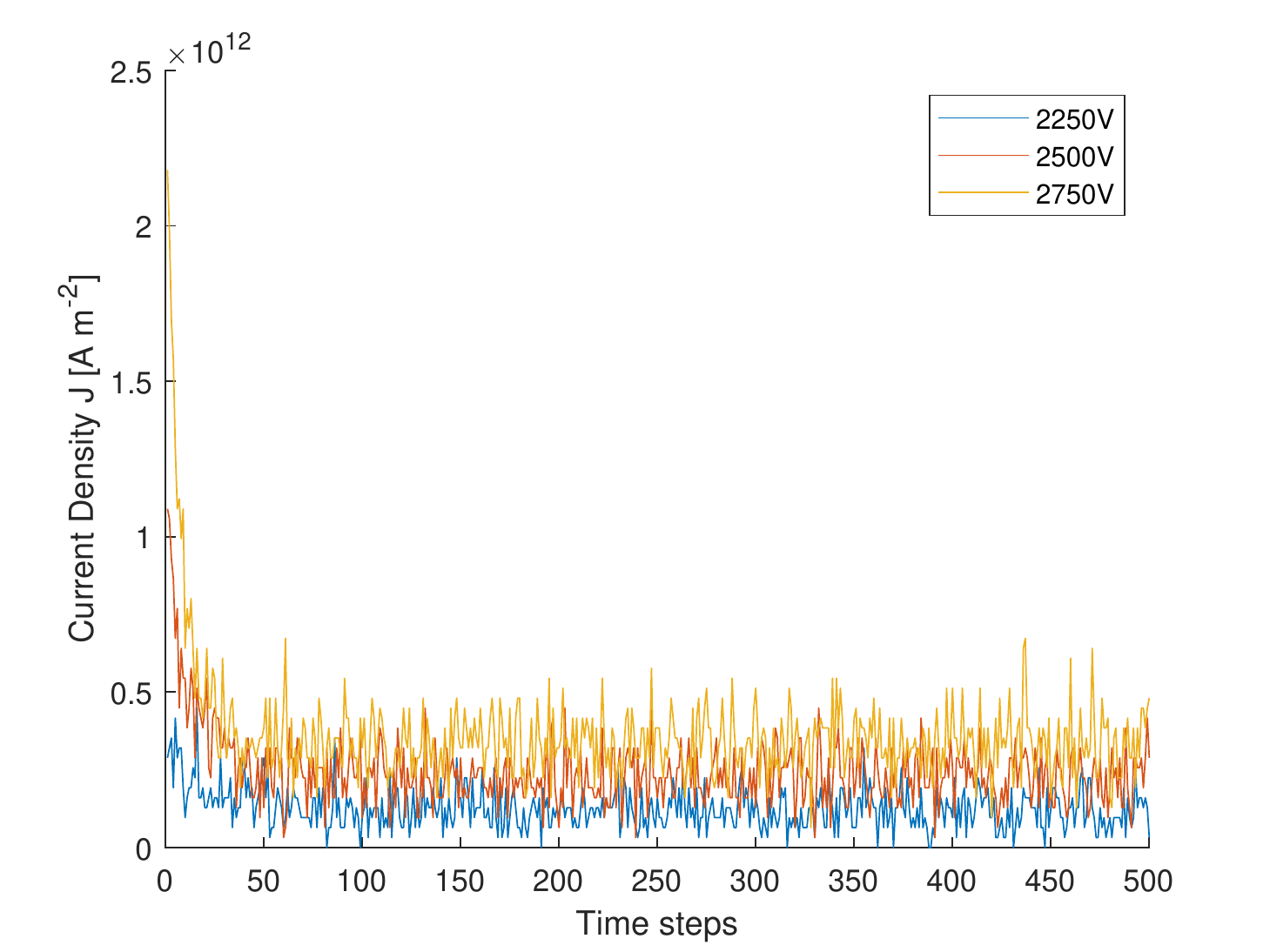}
    	\caption{Current density J [A/m$^2$] per time step in the simulation for the center emitter in the four emitter system. Gap spacing, $D$~=~1000~nm, gap voltage $V$~=~2250~-~2750~V, work function $\phi$~=~2.2~eV, side length $L$~=~100~nm, and emitter spacing $d$~=~210~nm.}
    	\label{fig:time_curr}
    \end{figure}

The current is calculated using the Ramo-Shockley theorem~\cite{ramo_currents_1939,doi:10.1063/1.1710367},
where we sum over the contribution from all electrons to the total current. The equation is
\begin{equation}
    i = q \sum \Vec{E} \cdot \Vec{v}\, , 
\end{equation}
where \(q\) and \(\vec{v}\) are the electron charge and its instantaneous velocity. The electric field \(\vec{E}\) is calculated at the electrons location with all other charges removed and the anode potential set to unity.
An example of simulated current is presented in Fig. \ref{fig:time_curr}. The initial pulse correspond to the first series of electrons hitting the anode.  In this simulation the number of electrons present in the gap varied between 4 and 68. During the simulations we keep track of the point of emission of individual electrons, and therefore it is possible to decompose the total current into components coming from each of the individual emitting patches.

The two models we describe in this paper have different simulation parameters. The two emitters model in~\cref{fig:model} uses a gap spacing of \(D = 2500\,\mathrm{nm}\), emitter side length of \(L = 100\,\mathrm{nm}\) and a work function equal to \(\phi = 4.7\,\mathrm{eV}\). The gap voltages examined are \(V = 25\,\mathrm{kV}\) and \(30\,\mathrm{kV}\) and the center-to-center spacing
(\(d\) in~\cref{fig:model}) is a free varying parameter.
The four emitter model seen in~\cref{fig:emitters} has a smaller gap spacing of \(D = 1000\,\mathrm{nm}\)
and a lower work function \(\phi = 2.2\,\mathrm{eV}\). Using a lower work function allows us to have a lower turn on voltage. The emitter side lengths studied are \(L = 50-125\,\mathrm{nm}\) in steps of \(25\,\mathrm{nm}\) and the gap voltages examined are \(V = 2250\,\mathrm{V}\), \(2500\,\mathrm{V}\) and \(2750\,\mathrm{V}\).
The free varying parameter is the middle-to-vertex distance (\(d\) in~\cref{fig:emitters}).

All of the simulations were run on a cluster at Reykjavik University with each run using a single node with 8 to 12 cores. The simulation time was usually between 6 and 24 hours depending on the number of emitters and parameters chosen.



\section{Results}

\begin{figure}[!t]
        	\centering
        	\subfloat[][Current density profile for the cathode.]{\label{fig:current_dens_a}\includegraphics[width=0.95\linewidth]{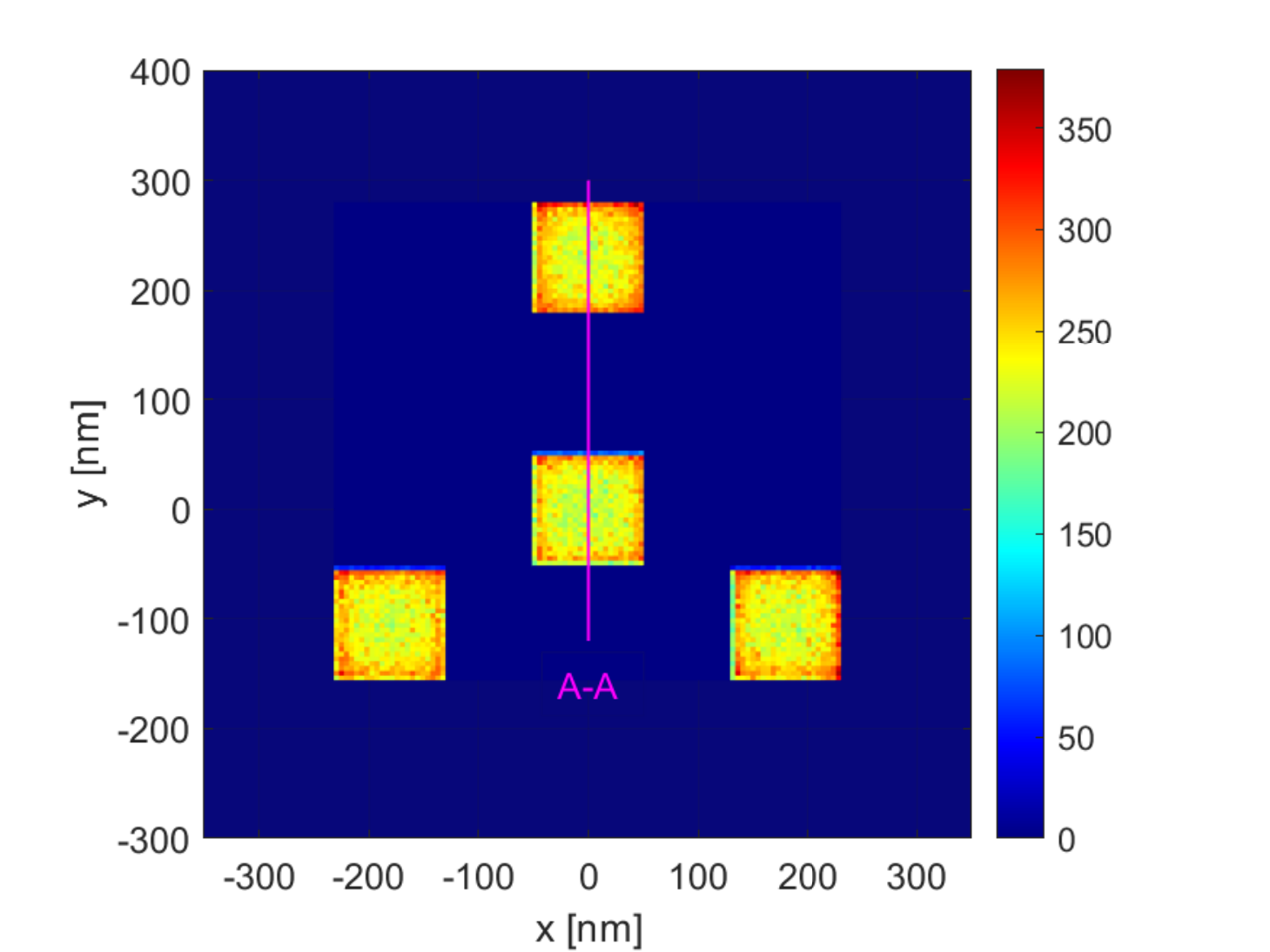}}%
        	\\
            \subfloat[][Current density profile for the anode.]{\label{fig:current_dens_b}\includegraphics[width=0.95\linewidth]{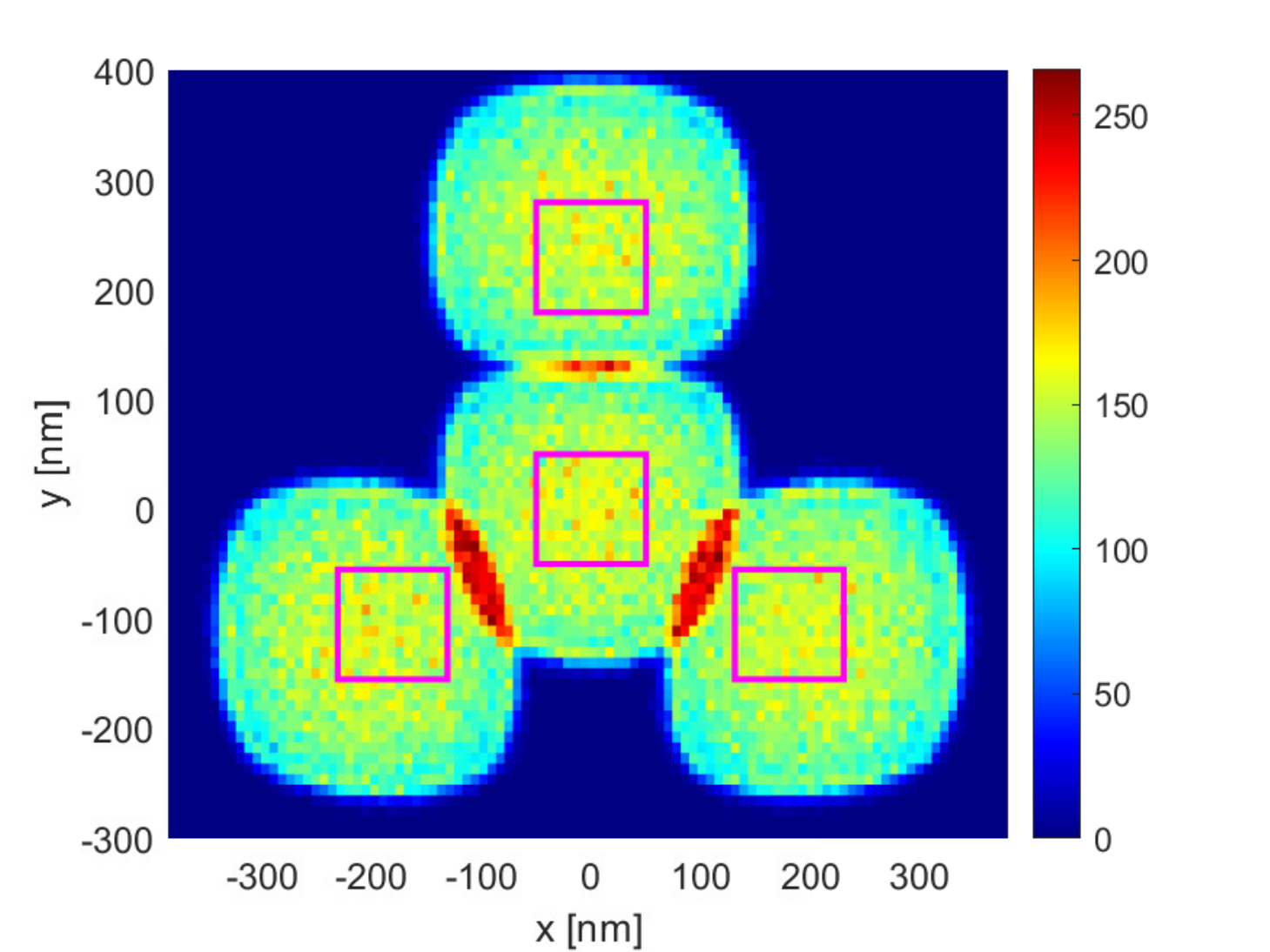}}%
            \\
            \subfloat[][Cross-sectional average of the current density for the cathode.]{\label{fig:current_dens_c}\includegraphics[width=0.95\linewidth]{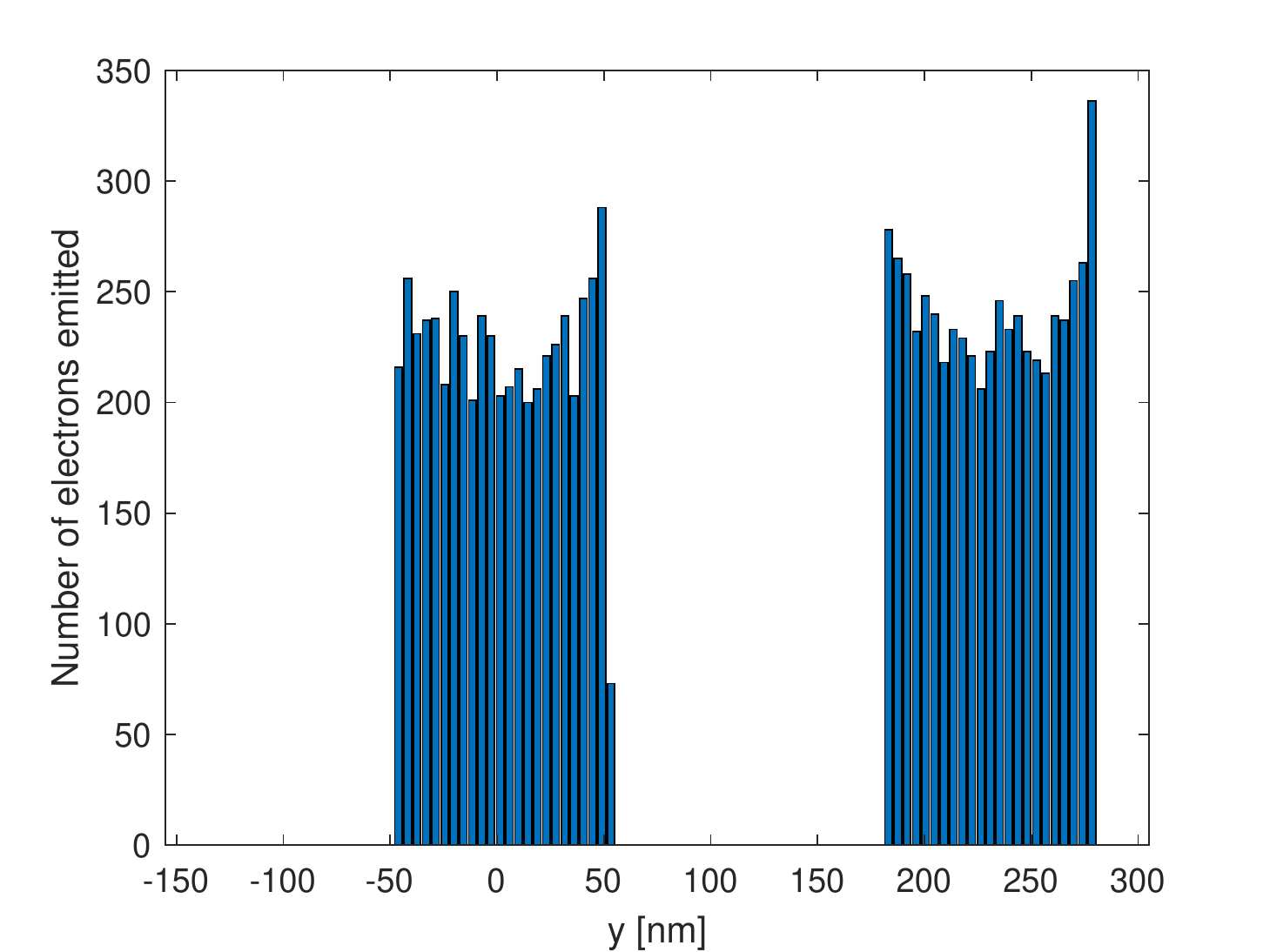}}%
        	\caption{Current density profiles in the (a) cathode, and (b) anode plane for the four emitter system. Gap spacing, $D$~=~1000~nm, gap voltage $V$~=~2750~V, work function $\phi$~=~2.2~eV, side length $L$~=~100~nm, and emitter spacing $d$~=~210~nm. Also shown in (c) is a cross-sectional average of the current density at the cathode taken along the cut shown in (a).}
        	\label{fig:current_dens}
        \end{figure}

It is illustrative to get a feel for the current distribution in the diode. \cref{fig:current_dens_a} shows a time integrated histogram of the point of emission of electrons from the cathode. As expected, the current density is highest at the edges of the emitting patches. \cref{fig:current_dens_b} shows a similar histogram taken in the plane of the anode. Here it can be seen that the beams emitted from the individual patches have spread out and become rounded (note the emitting patches on the cathode superimposed on the histogram). We also note that there are regions of higher density at the intersection of beams. Similar results are obtained for the two emitter system, although the symmetry is different. Finally, \cref{fig:current_dens_c} shows a histogram of the current density at the cathode for a section taken across the y-axis.

 We next look at the two emitter system for two different values of the gap voltage, 25~kV and 30~kV. We present the results in terms of the normalized emitter distance, $d/D$, and normalized current density, $J/J_{CL2D}$, where $J$ is the average current density from the emitters and
\begin{equation}\label{eq:2d_cl}
J_{CL2D}=J_{CL}\left (1+\frac{\sqrt{2}}{\pi L/D}  \right )    
\end{equation} 
is the two-dimensional Child-Langmuir current density for a square emitter of side-length $L$ in a diode with gap spacing $D$~\cite{koh_three-dimensional_2005}, where

\begin{equation}
\label{eq:cl}
    J_{CL}=\frac{4\epsilon _{0}}{9}\sqrt{\frac{2e}{m}}\frac{V^{3/2}}{D^2}
\end{equation}
is the well-known Child-Langmuir current density for a one-dimensional planar diode~\cite{PhysRevSeriesI.32.492,PhysRev.2.450} with $e$ being the fundamental charge, $m$ the electron mass, and $\epsilon_{0}$ the permittivity of vacuum.

\begin{figure}[!t]
        \centering
        \includegraphics[width=0.95\linewidth]{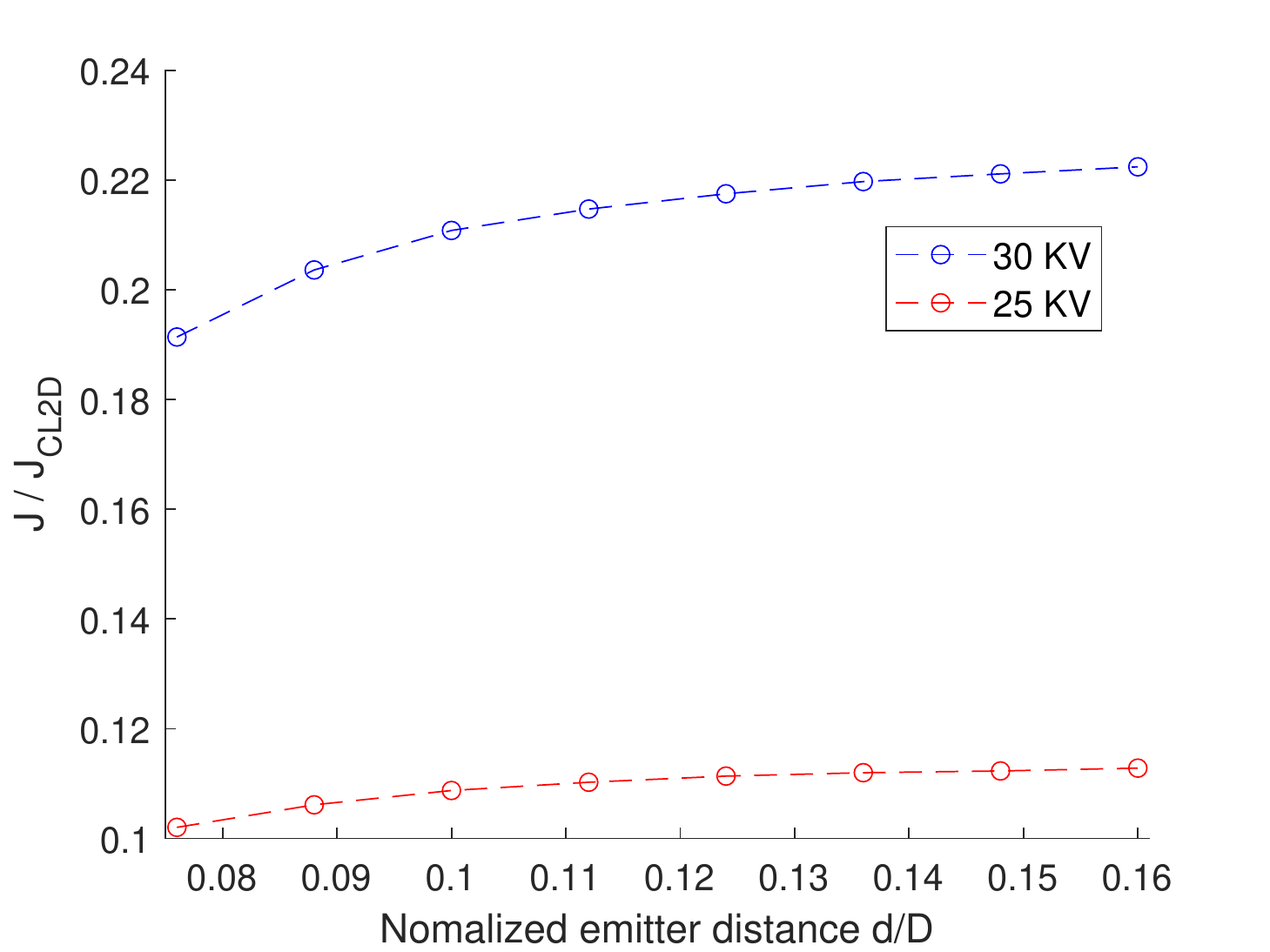}
    	\caption{Effect of emitter proximity in a two emitter system for two different gap voltages. The average current density is normalized by the current density as predicted by the 2D Child-Langmuir law. Here $L$~=~100~nm, $D$~=~2500~nm, work function $\phi$~=~4.7~eV, the emitter spacing, $d$, is varied. The blue curve is current density using 30~kV and the red curve is for 25~kV.}
    	\label{fig:emitt_prox-1}
    \end{figure}

In \cref{fig:emitt_prox-1} the blue curve shows results for a gap voltage of 30~kV, while the red shows the results for a gap voltage of 25~kV. We see that as the emitters are brought close to each other the average current density drops. This is readily understood as being due to the electrons emitted by one emitter reducing the surface electric field at the other emitter (due to the space-charge contribution), and thereby reducing emission. As the emitters are removed from each other their mutual interaction decreases and they act as two independent square emitters. We note that in both cases the normalized current density levels off as d increases. The general shape of the curve will be explained in context with the coupling parameter given in \cref{eq:coupling}. Regarding the effect of the gap voltage, we see that as the gap voltage is increased the normalized current also increases, which is in line with previous results~\cite{torfason_molecular_2015}. We also observe that the relative reduction in current, as the emitters are brought from infinite separation to being flush against each other, is greater in the high voltage case than in the low voltage case namely 14 percent and 10 percent respectively. This will be explained after results for the 4 emitter system have been presented. 
 
\begin{figure}[!t]
\centering
\includegraphics[width=0.95\linewidth]{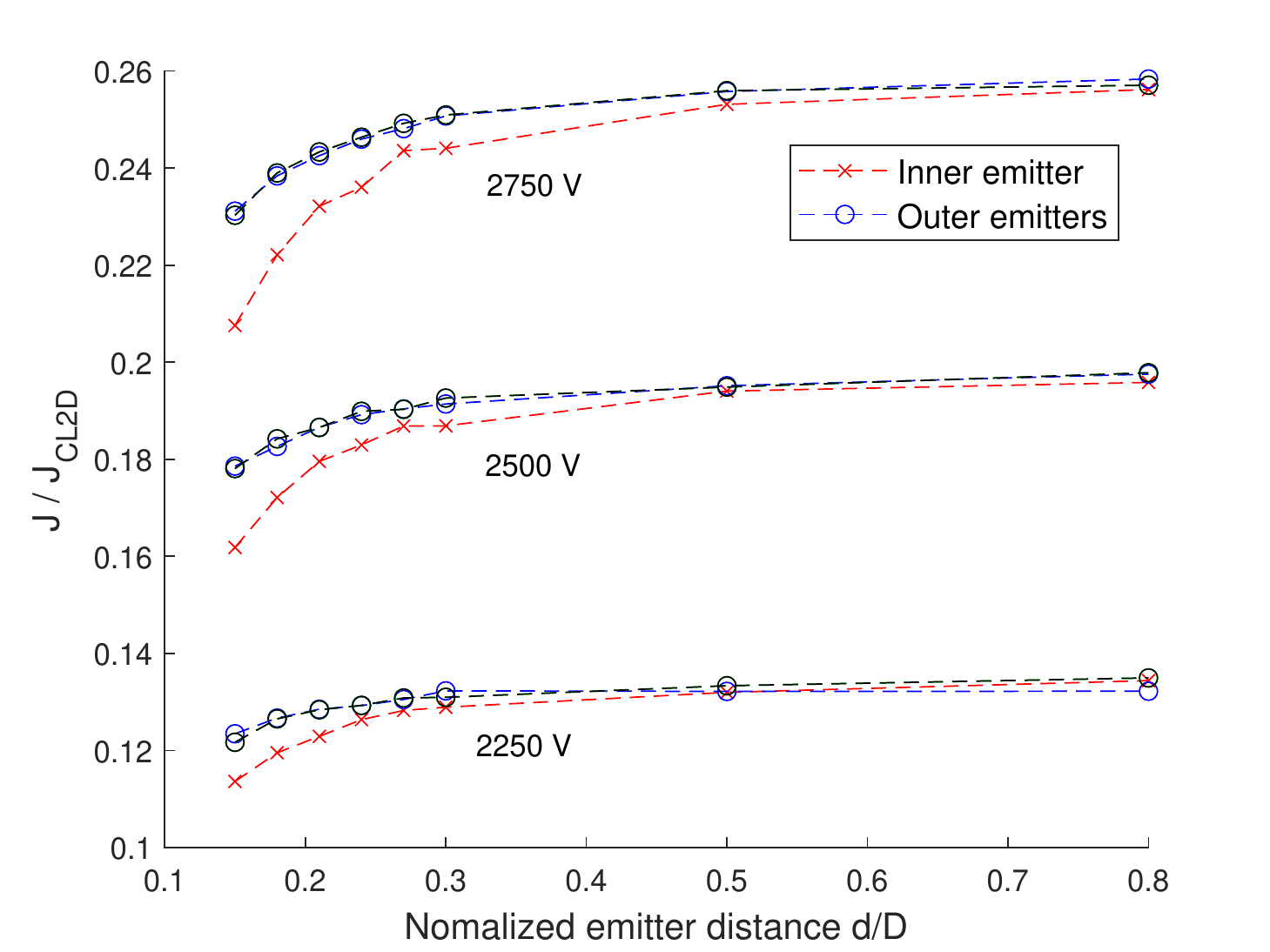}

\caption{Effect of emitter proximity in a four emitter system for 125~nm edge length and three different gap voltages. The average current density is normalized by the current density as predicted by the 2D Child-Langmuir law. Here $D$~=~1000~nm, work function $\phi$~=~2.2~eV, the emitter spacing, $d$, is varied. The gap voltage for the topmost, middle and bottom set of curves is 2750~V, 2500~V and 2250~V respectively. The red curves (x) show the normalized current density from the central emitter, and the black and blue curves (o) show the normalized current density from the outer emitters.}
        	\label{fig:emitt_prox-3}
\end{figure}

\begin{figure}[!t]
    \centering
    \includegraphics[width=0.95\linewidth]{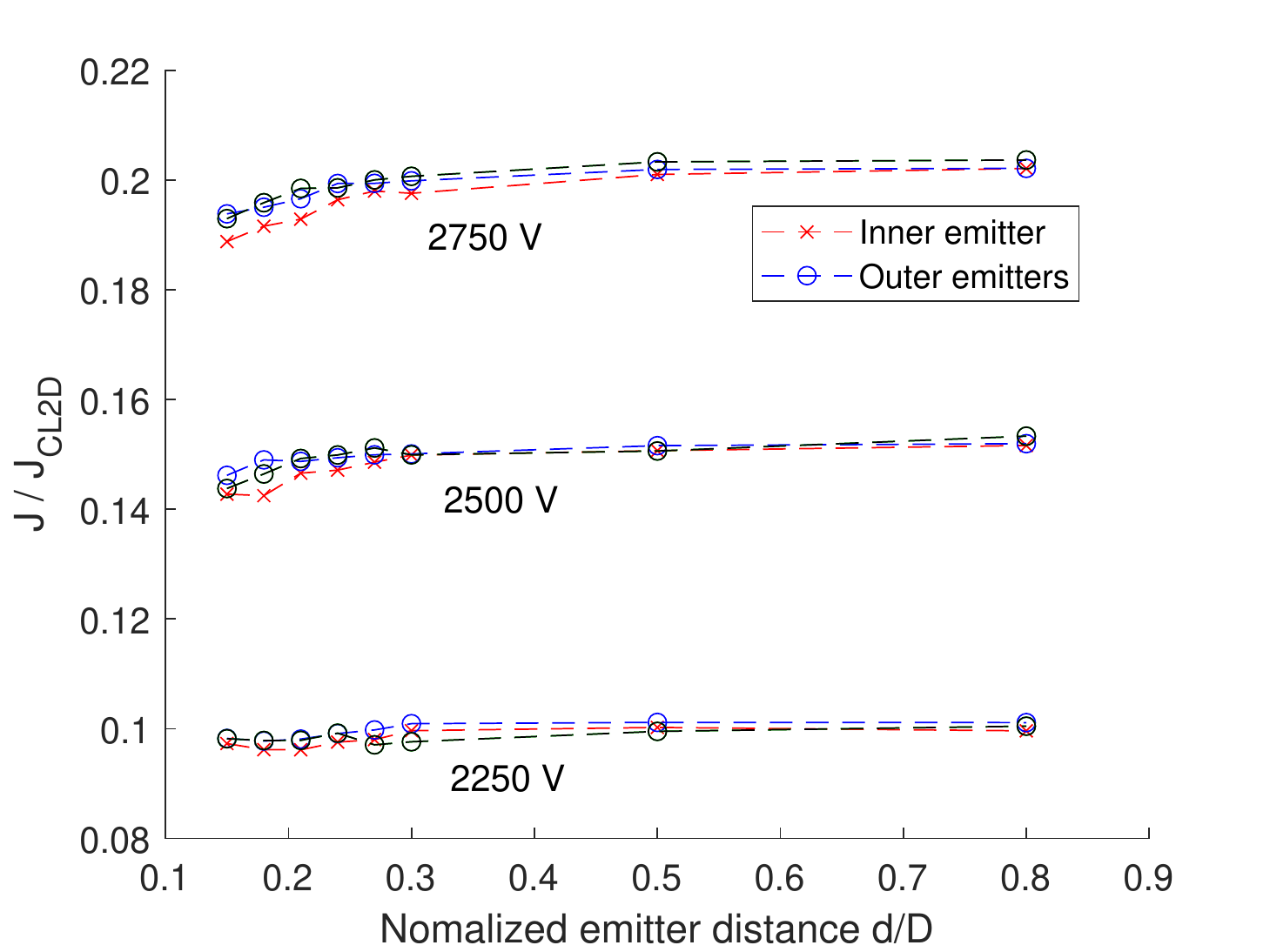}

    \caption{Effect of emitter proximity in a four emitter system for 75~nm edge length and three different gap voltages. The average current density is normalized by the current density as predicted by the 2D Child-Langmuir law. Here $D$~=~1000~nm, work function $\phi$~=~2.2~eV, the emitter spacing, $d$, is varied. The gap voltage for the topmost, middle and bottom set of curves is 2750~V, 2500~V and 2250~V respectively. The red curves (x) show the normalized current density from the central emitter, and the black and blue curves (o) show the normalized current density from the outer emitters.}
    \label{fig:emitt_prox-4}
    \end{figure}

Next we examine the four emitter system for six different parameter combinations. In all cases the gap spacing is 1000~nm and the work function of the emitters is 2.2~eV. For three cases the side length of the square emitters is 125~nm and for the other three the side length is 75~nm. For each of the sets of different side lengths we run a set of simulations at 2250~V, 2500~V, and 2750~V gap potential, where the emitter spacing is varied. \cref{fig:emitt_prox-3} shows the results of simulations with an emitter edge length of 125~nm, while \cref{fig:emitt_prox-4} shows the results for an edge length of 75~nm. Note that in these figures we look at the normalized current density from each of the emitters rather than averaged over all the emitters as was done in the two emitter model (where the emitters were mirror images of each other). We immediately note that in all cases the interior emitter is most strongly affected by its neighbors as it is surrounded by other emitters unlike those at the vertices. For the case of $L$~=~125~nm and $V$~=~2750~V the current from the central emitter is reduced by roughly 20 percent from its maximum value when the emitters are in closest proximity. We also note that the current density from the three outer emitters is nearly identical, even though the system does not quite possess perfect 120 degree rotational symmetry (as it would if the emitters were circular rather than square).

\begin{figure}[!t]
        \centering
        \includegraphics[width=0.95\linewidth]{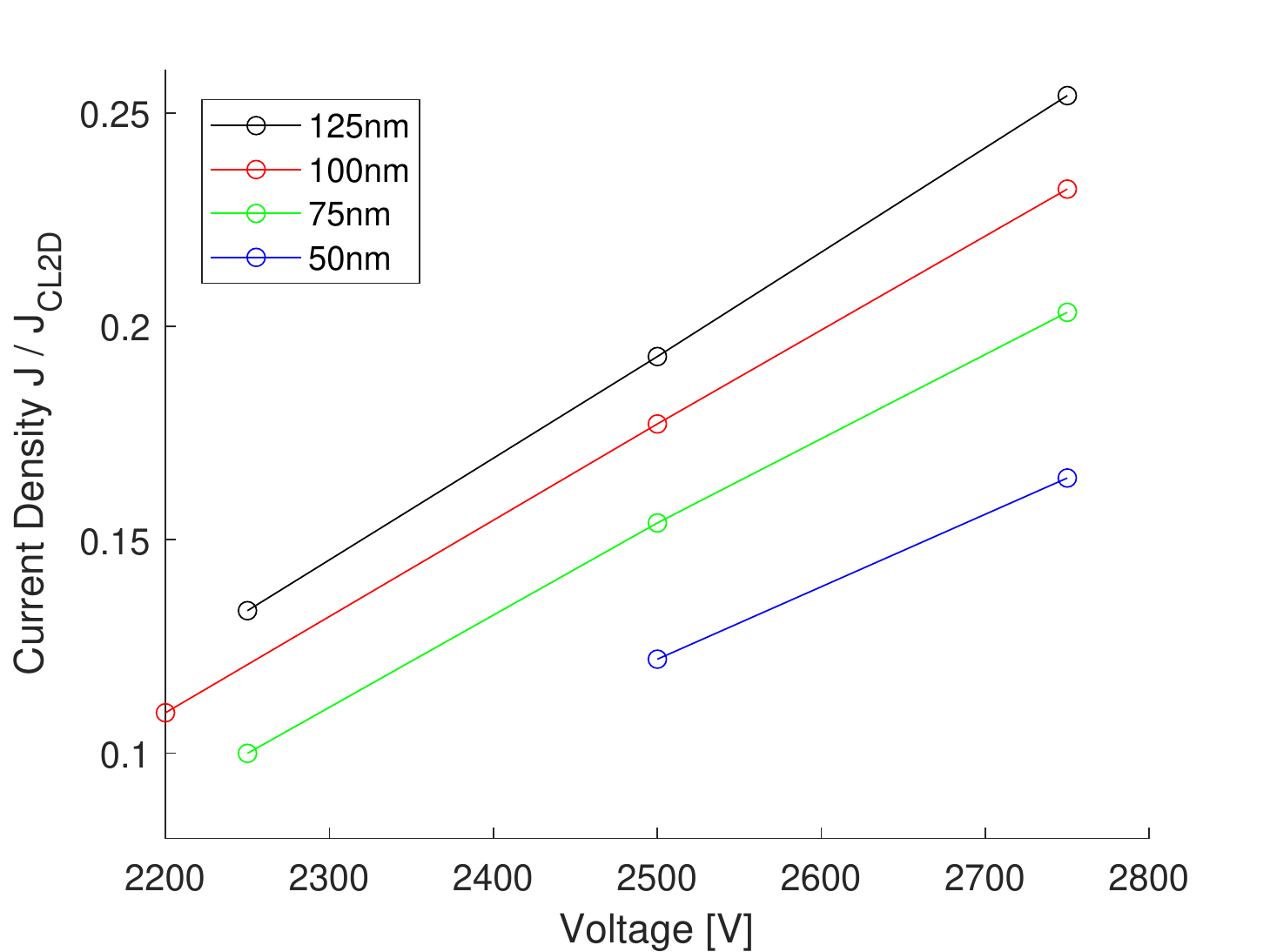}

        \caption{Effect of gap voltage and edge length on the average current density from the inner emitter in a four emitter model. Here $d$~=~300~nm, $D$~=~1000~nm and $\phi$~=~2.2~eV.}
    	\label{fig:gap_volt}
    \end{figure}

As can be seen from \cref{fig:emitt_prox-3,fig:emitt_prox-4,fig:gap_volt} the general trend is for the normalized current density to increase with applied voltage and emitter side length. This is in line with previous results for a single planar field emitter of finite area~\cite{torfason_molecular_2015}. It is interesting to compare the results for $L$~=~125~nm and $V$~=~2500~V to those obtained with $L$~=~75~nm and $V$~=~2750~V. The maximum value of the current density is comparable in both cases, but the minimum value is considerably lower for the former case than for the latter. The explanation is that, for the same current density, the total current coming from the larger emitter is 2.7 times higher than from the smaller emitter and thus the space-charge effect on the neighboring emitter is considerably stronger. In addition the spacing shown in the graph is the center-to-center spacing, $d/D$, but for equal values of $d/D$ the distance between the edges of neighboring emitters is smaller for the larger value of $L$.

To better understand the general shape of the curves displayed in \cref{fig:emitt_prox-1,fig:emitt_prox-3,fig:emitt_prox-4} we may look at a simple model of two emitters and how they interact. Consider two planar emitters, A and B, of equal size and shape located in a diode of gap length $D$ with some gap voltage $V$. The center-to center spacing is $d$. If we make the assumption that the center of charge of the electrons emitted from each of the emitters is located directly above the center of the emitter at a height of $D/\alpha$, then it is easy to find the ratio of the effect of the space charge from emitter A on the normal component of the surface electric field at the center of B, to the effect of the space charge from B on the normal component of the surface electric field at the center of B. This ratio is given by
\begin{equation}
\label{eq:coupling}
C=\frac{1}{\left ( 1+\left [ \alpha \frac{d}{D} \right ]^{2} \right )^{3/2}}
\end{equation}
where $C$ is the ratio of space charge effects described above~\cite{ilkov_synchronization_2015}, and $\alpha$ ranges from 3 for a vanishingly small current density to 4 for Child-Langmuir current density. In either case, for $d/D >$ ~0.5 the value of C drops below 0.2, meaning that the influence of the space charge from emitter A on the surface field of B is less than 20 percent of the effect of space charge from emitter B on the surface field. On the other hand if $d/D <$~0.1~C will range from \mbox{0.8~-~0.9}. This geometric effect explains the general shape of the curves in \cref{fig:emitt_prox-1,fig:emitt_prox-3,fig:emitt_prox-4} in the sense that the mutual space charge effect drops rather rapidly with $d/D$. However, the degree of reduction of the current from an emitter, due to space-charge effects from its neighbors, is related to the work function, applied voltage and emitter dimensions in a more complex way. For instance, we observe in \cref{fig:emitt_prox-3} a more pronounced relative decrease in current density with smaller spacing for higher voltage, due to the fact that the total current is higher.  A simple example of how material properties can influence how the emitted current is affected by space charge follows.

\begin{figure}[!t]
        \centering
        \includegraphics[width=0.95\linewidth]{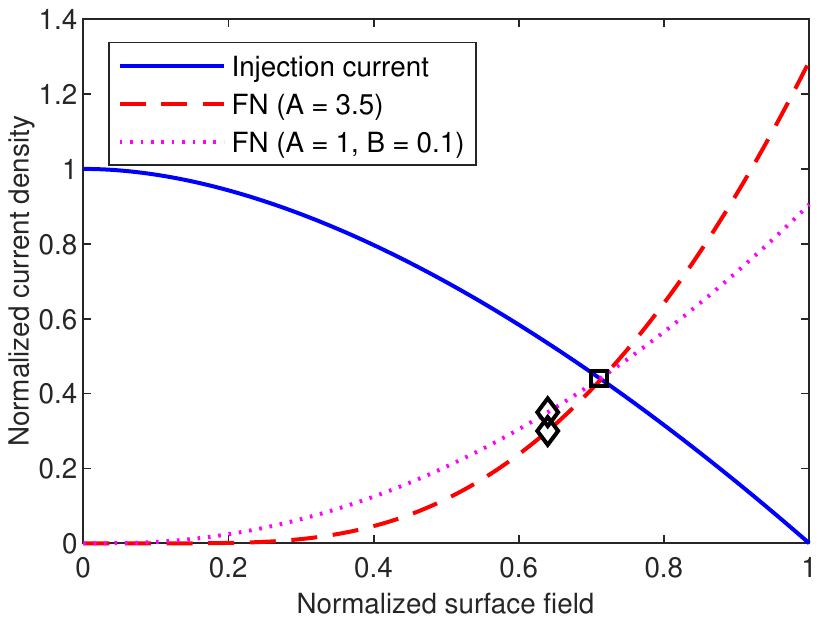}

        \caption{Normalized surface electric field versus injection current density, and normalized field emission current density calculated from~\cref{eq:fn_norm}. The solid blue line depicts the normalized injection current density from~\cref{eq:j_inj_1,eq:j_inj_2}. The dashed red line depicts the normalized Fowler-Nordheim current density for $A=3.5$ and $B=1$, whereas the dotted magenta line depicts the normalized Fowler-Nordheim current density for $A=1$ and $B=0.1$. The square depicts the equilibrium current density for field emission in a one dimensional diode, and the diamonds depict the reduced current density if an external charge were to reduce the surface electric field by 10 percent.}
    	\label{fig:surface_elec}
\end{figure}

Consider a one-dimensional diode with gap spacing $D$ and applied voltage $V$. Electrons are emitted uniformly from the cathode with negligible emission velocity and current density $J_{inj}$ then by solving Poisson's equation with energy and flux conservation it is a fairly straightforward exercise to show that the relation between the injected current and the magnitude of the surface electric field at the cathode is given by:
\begin{equation}\label{eq:j_inj_1}
    \overline{J}_{inj}=\frac{1}{2}\left ( 1 + \sqrt{1+\frac{27}{4}{\overline{E}_{0}}^{2}\left ( \overline{E}_{0}-1 \right )} \right )
\end{equation}
for $\overline{E}_{0} <$ 2/3, and
\begin{equation}\label{eq:j_inj_2}
    \overline{J}_{inj}=\frac{1}{2}\left ( 1 - \sqrt{1+\frac{27}{4}{\overline{E}_{0}}^{2}\left ( \overline{E}_{0}-1 \right )} \right )
\end{equation}
for $\overline{E}_{0} >$ 2/3. Here $\overline{J}_{inj}$ is the normalized injection current density and $\overline{E}_{0}$ is the normalized electric field at the cathode surface. The normalization factors for $J_{inj}$ and the cathode electric field are the Child-Langmuir current density, given by~\cref{eq:cl}, and the vacuum electric field $V/D$ respectively. A simplified form of the Fowler-Nordheim equation in normalized form can then be written as~\cite{lau_electron-emission_1994}:
\begin{equation}\label{eq:fn_norm}
    \overline{J}_{FN}=A\overline{E}_{0}^{2}exp(-B/\overline{E}_{0})
\end{equation}
where $A$ and $B$ are constants dependant on the work function and other parameters. The equilibrium value for field emitted current density and surface electric field are then found from the intersection of the curves for $\overline{J}_{inj}$ and $\overline{J}_{FN}$. \cref{fig:surface_elec} shows curves for two different types of field emitters (with different material properties) where the parameters $A$ and $B$ are chosen so that the equilibrium current is the same, but the shape of the curves differs. If some external charge were to depress the surface electric field further  (which is analogous to the effect on field emission from an emitter due to its neighbors) the value of $\overline{J}_{FN}$ would decrease accordingly. This can be seen in \cref{fig:surface_elec}, where the original equilibrium current density value for either of the two field emitters with different values of $A$ and $B$ is marked with a square and the current density due to a 10 percent reduction of the surface electric field marked with a diamond. We note that for the steeper Fowler-Nordheim curve a 10 percent reduction in surface field results in a 30 percent reduction in the current density whereas it results in a 20 percent reduction in the current density for the flatter Fowler-Nordheim curve. This explains qualitatively why the maximum of the mutual space-charge effect seen in \cref{fig:emitt_prox-1,fig:emitt_prox-3,fig:emitt_prox-4}, is affected by parameters other than the spacing.
\section{Conclusion}
In this work we have examined mutual space-charge effects between neighboring field emitters of finite extent in a planar diode. Simulations indicate that these effects can decrease the current density from an emitter by 10 - 20 percent of the value that would be expected if it were not influenced by the current streaming from its neighbors. Simulations indicate that in the case of emitters with a side length of roughly 100~nm in a diode with gap spacing on the order of 1000 - 2000~nm the emitters become essentially independent of each other when the spacing between them lies in the range of $D/10$ to $D/2$, though this is quite dependent on emitter size, work function etc. This indicates that, in closely spaced field emitter arrays where the pitch is much smaller than the diode gap, mutual space-charge effects will be of importance. Our model also indicates that field emitters in the array are much more strongly affected than those at the edge of the array.

It is of interest to compare the results from this study to those of Harris et al.~\cite{Harris_2015} where the effects of spacing between ungated carbon fiber like emitters is considered. For the fiber arrays there are two effects at play: Screening of the applied field due to the geometry of the array, and space-charge effect from emitted current influencing the field at the point of emission. Although the characteristic size of the fiber arrays is on the order of 1 microns for the fiber tip radius to 1000 microns for fiber length, as compared to the submicron scale of the system under study in this paper, some conclusions can be drawn. In Harris' paper, the predominant effect is screening which can change the field enhancement by well over 10 percent depending on the tip radius and spacing between the fibers. This would presumably outweigh the effect of mutual space-charge interaction between emitters due to the strong dependence of the current on the surface field at the emission site. For emission sites with smaller aspect ratios, or in the case of high current and closely spaced emitters, the mutual space charge effect could be of greater importance, but that is subject to further investigation.

The model presented in this paper is rather simple but is illustrative of the basic physics governing mutual space-charge effects as screening due to the proximity of separate emitters does not affect emission as it would if the emitters were the typical tips. Nonetheless, to better model practical FEA's, simulations on mutual space-charge effects between neighboring field emitting tips are in preparation.   


%



\section*{Acknowledgment}

This material is based upon work supported by the Air Force Office of Scientific Research under award number FA9550-18-1-7011, and by the Icelandic Research Fund grant number 174127-051.
Any opinions, findings, and conclusions or recommendations expressed in this material are those of the author and do not necessarily reflect the views of the United States Air Force.

\ifCLASSOPTIONcaptionsoff
  \newpage
\fi



%
\printbibliography

%
\begin{IEEEbiography}[{\includegraphics[width=1in,height=1.25in,clip,keepaspectratio]{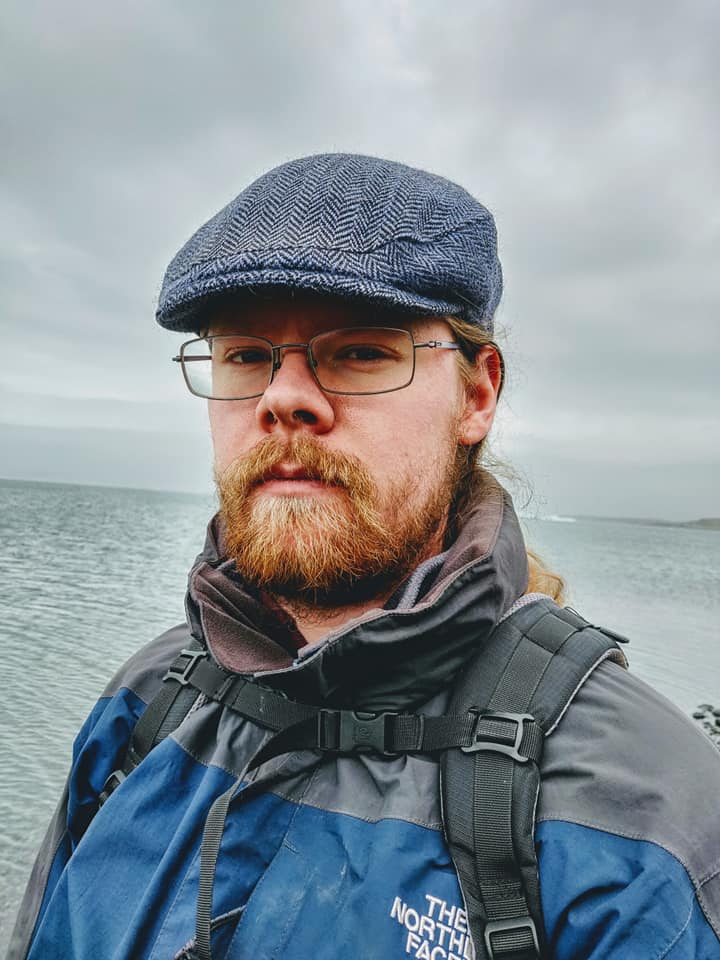}}]{Hákon Valur Haraldsson}
Hákon received a BSc degree in applied electrical engineering from the University of Reykjavik in 2017 and is currently working on an MSc degree in electrical energy engineering at the same university. He did his bachelors thesis on Field emission in micro vacuum devices and his current master thesis in on large scale simulations of the Icelandic power system in relation to electric vehicles. He is also a master electrician and has worked as an engineer designing electrical systems for various projects in Iceland. Current research interests are in electron emission modelling, impact of electric vehicles and power system simulation. 
\end{IEEEbiography}

\begin{IEEEbiography}
[{\includegraphics[width=1in,height=1.25in,clip,keepaspectratio]{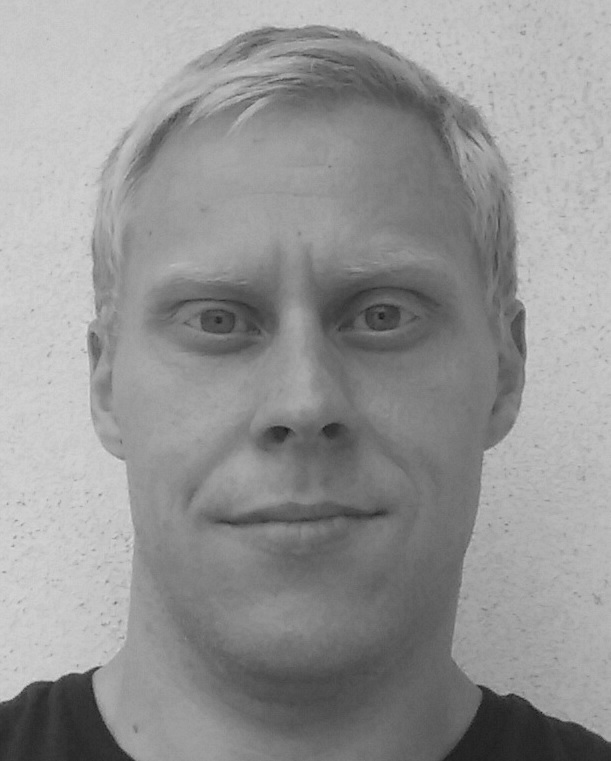}}]
{Kristinn Torfason}
Kristinn Torfason was born in Reykjavík, Iceland in 1984. He received a BSc in 
Physics from the University of Iceland in 2007 and a masters degree in 2009 in the 
same field. In 2013 he finished a joint degree in computational physics from the University 
of Iceland and Reykjavik University. The same year he joined the Department of Engineering at Reykjavik University as a Postdoctoral Research Associate. His current research
interests are vacuum electronics and quantum transport.
\end{IEEEbiography}

\begin{IEEEbiography}
[{\includegraphics[width=1in,height=1.25in,clip,keepaspectratio]{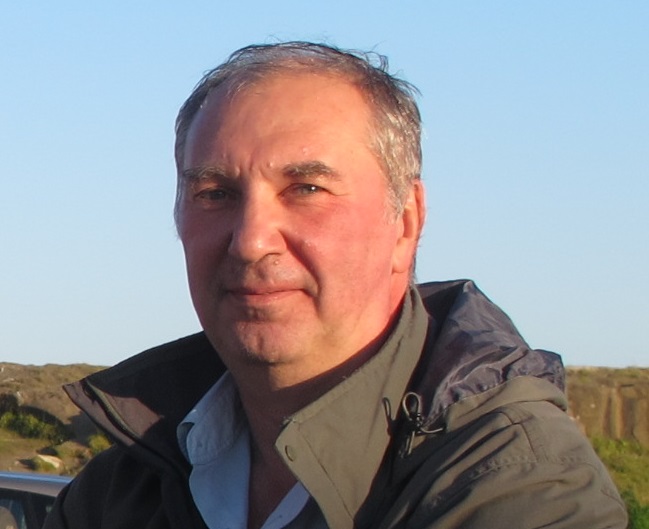}}]
{Andrei Manolescu}
Andrei Manolescu, born in Bucharest, Romania in 1958, studied physics at University of Bucharest and received a doctor degree in physics at the Institute of Atomic Physics Bucharest in 1992 where
he worked until 1999. Between 1999-2007 he did research in human genetics at Decode Genetics Iceland. He is now a Professor at Reykjavik University where he moved in 2008. He is doing theoretical
research on transport properties of nanoelectronic devices using modelling and numerical calculations. 
\end{IEEEbiography}

\begin{IEEEbiography}[{\includegraphics[width=1in,height=1.25in,clip,keepaspectratio]{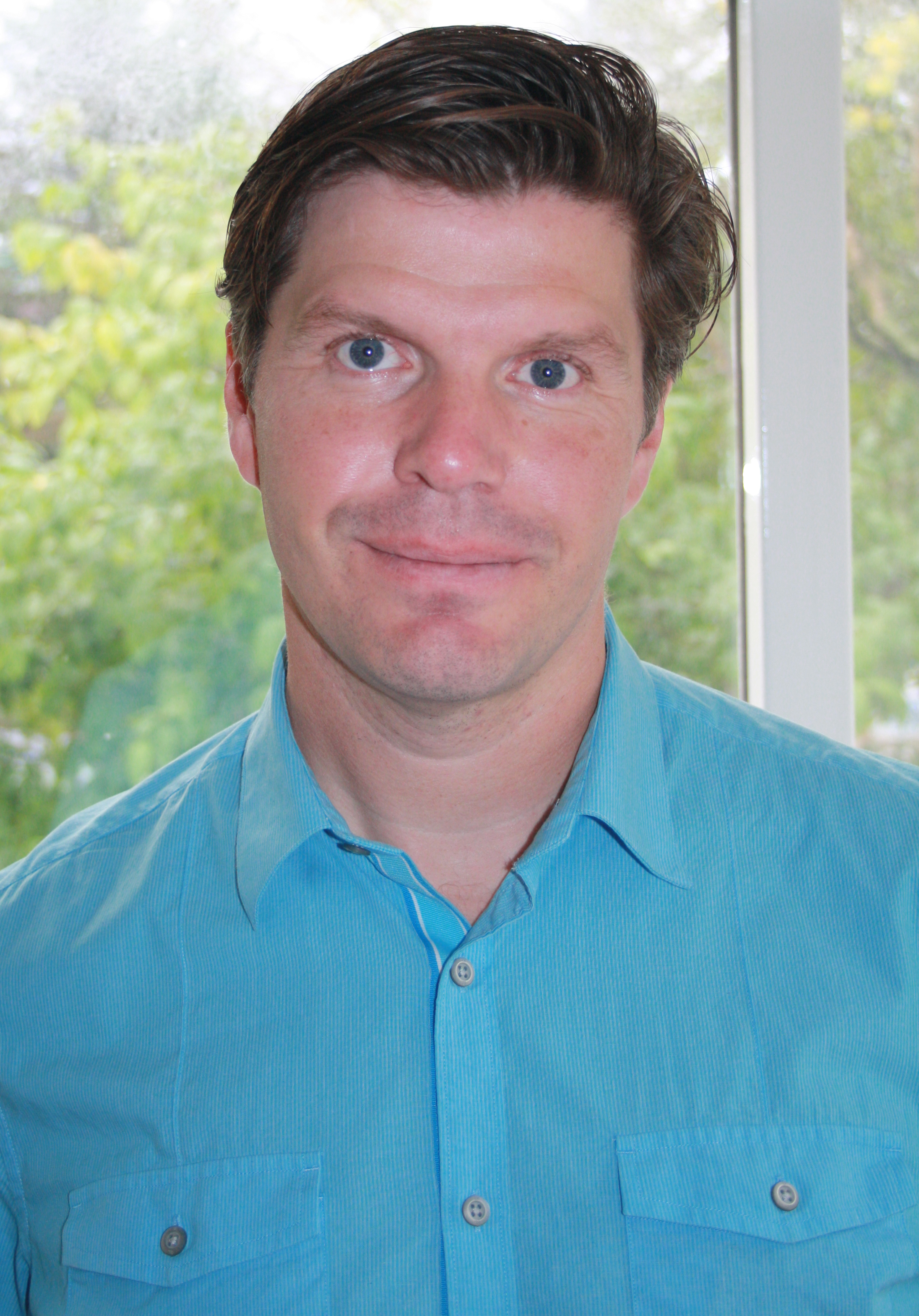}}]{Ágúst Valfells}

received the Ph.D. degree from the University of Michigan, Ann Arbor, MI, USA, in 2000. His thesis work was on multipactor discharge. From 2000 to 2003 he was a Research Associate at the Institute for Research in Electronics and Applied Physics at the University of Maryland, where he worked on the University of Maryland Electron Ring and modelling of secondary electron emission for a library of codes for depressed collector design. Subsequently he worked at the National Energy Authority of Iceland, where he was responsible for a platform to reduce fossil fuel use in the transport. Since 2005 he has been with the Department of Engineering at Reykjavík University, where he is a Professor. His current research interests are in diode physics, geothermal reservoir modelling and electron emission modelling.
\end{IEEEbiography}





\end{document}